\documentclass[rmp, 
aps, 
twocolumn, 
amsmath, 
amssymb,
superscriptaddress
]{revtex4-2}

\usepackage{graphicx}
\usepackage{amsfonts}
\usepackage{natbib}
\usepackage{bm}
\usepackage{hyperref}
\usepackage{dsfont}

\usepackage[usenames,dvipsnames]{xcolor}
\DeclareMathOperator*{\argmax}{\arg\max}

\setcitestyle{numbers,square}

\begin{document}
\title{Concise network models of memory dynamics\\ reveal explainable patterns in path data}

\author{Rohit Sahasrabuddhe}
\email{rohit.sahasrabuddhe@maths.ox.ac.uk}
\affiliation{Mathematical Institute, University of Oxford, Oxford, UK}
\affiliation{Institute for New Economic Thinking, University of Oxford, Oxford, UK}

\author{Renaud Lambiotte}
%\email{renaud.lambiotte@maths.ox.ac.uk}
\affiliation{Mathematical Institute, University of Oxford, Oxford, UK}

\author{Martin Rosvall}
%\email{martin.rosvall@umu.se}
\affiliation{Integrated Science Lab, Department of Physics, Ume\aa\; University, Ume\aa, Sweden\looseness=-1}

\date{\today}

\begin{abstract}

Networks are a powerful tool to model the structure and dynamics of complex systems across scales. 
Direct connections between system components are often represented as edges, while paths and walks capture indirect interactions.
This approach assumes that flows in the system are sequences of independent transitions.
Path data from real-world systems often have higher-order dependencies, which require more sophisticated models.
In this work, we propose a method to construct concise networks from path data that interpolate between first and second-order models.
We prioritise simplicity and interpretability by creating \textit{state nodes} that capture latent modes of second-order effects and introducing an interpretable measure to balance model size and accuracy.
In both synthetic and real-world applications, our method reveals large-scale memory patterns and constructs concise networks that provide insights beyond the first-order model at the fraction of the size of a second-order model.
\end{abstract}
\maketitle

\section*{Introduction}
Networks offer a versatile framework for modelling complex systems, abstracting their components as nodes and interactions as edges \cite{newman2018networks}. These edges often represent \textit{flows} of some quantity, such as passengers between airports, information between people, or workers between occupations. While edges capture only direct flows between pairs of nodes, a key feature of networks is their ability to model indirect flows as paths or walks. This capacity allows researchers to analyse systems across scales \cite{lambiotte2021modularity}, revealing meso-scale structures such as communities and roles, and macro-scale patterns such as hierarchies and rankings. Capturing these patterns relies on assuming that flows are transitive -- given flow from node $i$ to node $j$ and from node $j$ to node $k$, there is implied indirect flow from $i$ to $k$ through $j$, often modelled as a first-order Markov process.

To validate this Markovian assumption, researchers analyse empirical squence or trajectory data. Consider, sequences of the form $(x_1 \rightarrow x_2 \rightarrow \cdots \rightarrow x_l)$, where $x_t$ are system components such as airports in itineraries or occupations in careers. In a first-order network, nodes represent these entities and edges capture first-order transition rates. This memoryless model treats the flow $i \rightarrow j \rightarrow k$ as two independent steps, with a combined probability $P(x_{t+1}=j \, | \, x_t=i) P(x_{t+1}=k \, | \, x_t=j)$. But many real-world systems often have higher-order dependencies \cite{chierichetti2012web, kareiva1983analyzing, meiss2008ranking, west2012human} that memoryless network models fail to capture. To address this issue, researchers have introduced network models with memory \cite{butts2009revisiting, rosvall2014memory, xu2016representing, scholtes2017network, lambiotte2019networks}, expanding the network toolbox to applications where the Markovian assumption is too restrictive. A natural extension is a second-order network, which incorporates one-step memory with \textit{state nodes} of the form $j|i$ -- indicating arrival at $j$ from $i$, where $j$ and $i$ correspond to \textit{physical nodes} -- the real-world system components. Edges capture the transition rate from $j|i$ to $k|j$ as $P(x_{t+1}=k \, | \, x_t=j, \, x_{t-1}=i)$. This approach can be generalised to build networks with \textit{fixed-order} memory of any order, capturing increasingly complex dependencies.
 
Combining Markov models up to a maximum order in a multi-order approach significantly improves next-element prediction in real-world systems \cite{scholtes2017network, gote2023predicting}. However, modelling an entire system with a single fixed Markov order imposes a strong constraint, as real systems may exhibit dependencies spanning multiple orders. The heterogeneous distribution of observations in real data can cause such models to simultaneously overfit and underfit different parts of the system. Higher-order networks with state nodes used only where required have been proposed to overcome these challenges \cite{xu2016representing, saebi2020efficient}. While these approaches mitigate overfitting, they rely on heuristics to estimate the importance of memory effects, limiting their interpretability and scalability for even moderately sized systems. Effective models must balance simplicity and predictive accuracy \cite{queiros2022toward}.

In this work, we focus on building simple and interpretable models that interpolate between first- and second-order networks. Using non-negative matrix factorisation, we construct state nodes that represent latent \textit{modes}, capturing prominent patterns in second-order dynamics. We introduce a simple measure of model performance to balance quality and complexity, and incorporate a Bayesian prior to mitigate overfitting. We processes each physical node independently, enabling parallel execution. After validating our method on synthetic data, we apply it to two real-world systems: air travel and information spread. In both cases, the method captures critical memory effects using a minimal number of interpretable state nodes, providing insights beyond those offered by first-order models.

\section*{Results}

\subsection*{Concise network models of path data}
\begin{figure*}[ht]
    \centering
    \includegraphics[width=.9\linewidth]{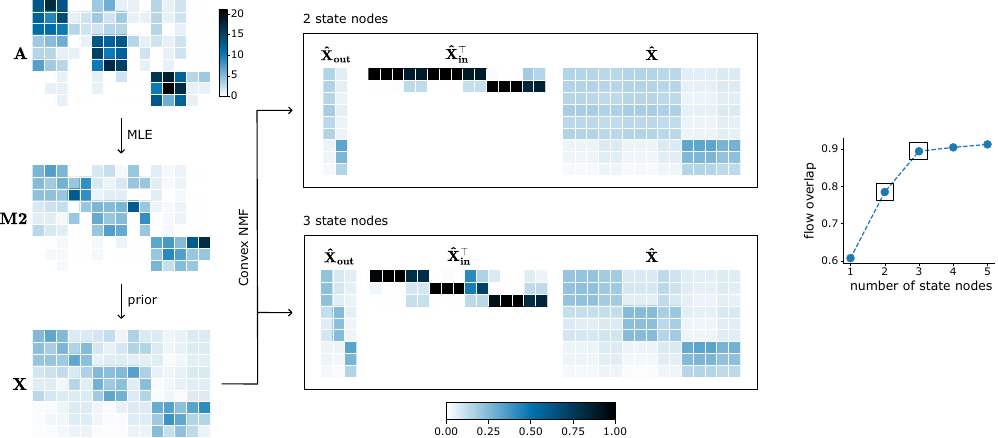}
    \caption{\textbf{Method schematic.} We illustrate the steps in our method with an example of a physical node. The matrix of observation counts, $\mathbf{A}$ (top left), has a block structure with 3 equally sized blocks each of its 15 predecessors and 9 successors. While most of the flow is between the corresponding blocks, the first and second blocks have more transitions between them than with the third. 2 predecessors in each block are undersampled. The MLE of second-order transition rates, $\mathbf{M2}$ (centre left), overfits to the undersampled nodes, which is regularised by a Bayesian prior to obtain $\mathbf{X}$ (bottom left). We find representations with 2 and 3 state nodes (centre) that capture the planted behaviour. Minimal gain in flow overlap for more state nodes (centre right) shows that we capture all the large-scale patterns with 3 state nodes. All matrices except $\mathbf{A}$ share a colourbar (bottom centre).}
    \label{fig:methods_schematic}
\end{figure*}
Understanding the dynamics of complex systems requires models that balance simplicity and explanatory power. We focus on constructing concise and interpretable network representations that capture essential memory effects. As a first step, we create state nodes independently for each physical node $j$ following the steps illustrated with an example in Fig.~\ref{fig:methods_schematic}. Let $\mathbf{A}$ be the matrix of observation counts of the trigrams through $j$, where $A_{ki}$ is the frequency of $i \rightarrow j \rightarrow k$. We use $i$ and $k$ to refer to predecessors and successors respectively\footnote{The sets of predecessors and successors may overlap without any effect.}. In the example, $|i|=15$, $|k|=9$, and $\mathbf{A}$ has a block structure with 3 equally sized blocks each of predecessors and successors. While most transitions are between corresponding blocks, the first and second blocks have more flow between them than with the third. Two predecessors in each block are undersampled. Let matrix $\mathbf{M2}$ be the Maximum Likelihood Estimate (MLE) of second-order transition rates
\begin{equation}
    M2_{ki} := \frac{A_{ki}}{\sum_{k'}A_{k'i}}.
\end{equation}
$\mathbf{M2}$ is susceptible to overfitting (see example), which we combat by regularising it with a Bayesian prior.

\paragraph*{Prior.} Let vector $\mathbf{M1}$ be the MLE of the first-order transition rates
\begin{equation}
    M1_{k} := \frac{\sum_{i'} A_{ki'}}{\sum_{k', i'}A_{k'i'}}.
\end{equation}
We introduce a Dirichlet prior with parameters $\mu \mathbf{M1}$ $\left(\mu \in \mathbb{R}^+\right)$ to each column $\mathbf{M2_{\cdot i}}$, which helps regularise under-sampled predecessors and reduce overfitting. This approach corresponds to assigning a pseudocount of $\mu M1_k$ to $A_{ki}$. Let matrix $\mathbf{X}$ be the posterior mean of the second-order transition rates
\begin{equation}
     X_{ki} := \frac{A_{ki} + \mu M1_k}{\sum_{k'}A_{k'i} + \mu}.
     \label{eq:Xki}
\end{equation}
We can write the column $\mathbf{X_{\cdot i}}$ as
\begin{equation}
     \mathbf{X_{\cdot i}} = \left(\frac{\mu}{\mu + n_i}\right) \mathbf{M1} + \left(1- \frac{\mu}{\mu + n_i}\right) \mathbf{M2_{\cdot i}},
     \label{eq:Xi}
\end{equation}
where $n_i = \sum_{k'}A_{k'i}$. In cases where system knowledge cannot inform the choice of $\mu$, we use leave-one-out cross-validation (see Methods). The prior is stronger for undersampled predecessors, regularising their transition rates (see Fig.~\ref{fig:methods_schematic}).

\paragraph*{Constructing state nodes.} 
We aim to represent the target dynamics in $\mathbf{X}$ by constructing interpretable state nodes that capture prominent second-order patterns in the transition rates to successors.
Specifically, we estimate transition rates $\hat{P}(i \rightarrow \alpha)$ from $i$ to state node $\alpha$ and $\hat{P}(\alpha \rightarrow k)$ from $\alpha$ to $k$ and approximate
\begin{equation}
    X_{ki} \approx \hat{X}_{ki} := \sum_\alpha \hat{P}(i \rightarrow \alpha) \hat{P}(\alpha \rightarrow k).
\end{equation}
We can rewrite this expression as 
\begin{align}
    \mathbf{\hat{X}} &= \mathbf{\hat{X}_\text{out}} \mathbf{\hat{X}_\text{in}^\top}, \text{ where} \nonumber\\
    (\hat{X}_\text{in})_{i\alpha} &:= \hat{P}(i \rightarrow \alpha), \nonumber\\
    (\hat{X}_\text{out})_{k\alpha} &:= \hat{P}(\alpha \rightarrow k).
    \label{eq:Xhat}
\end{align}
Given the number of state nodes $r \ll \min(|i|, |k|)$, we use Convex Non-negative Matrix Factorisation to find the optimum rank-$r$ approximation such that $\mathbf{\hat{X}_\text{in}}$ and $\mathbf{\hat{X}_\text{out}}$ are themselves transition matrices (see Methods). A key feature of this step is that each column of $\mathbf{\hat{X}_\text{out}}$ is a convex combination of the columns of $\mathbf{X}$, creating state nodes that capture plausible behaviour. For instance, at rank 2 in our example (Fig.~\ref{fig:methods_schematic}), we identify the coarser behaviour -- the first two blocks of successors are merged in $\mathbf{\hat{X}_\text{out}}$. Adding another state node separates the first and second blocks, recovering the planted dynamics. In both cases, the well-sampled predecessors are assigned exclusively to a single state node each in $\mathbf{\hat{X}_\text{in}}$. The uncertainty in the behaviour of the undersampled ones leads to them being modelled as a mixture of behaviours.

For each physical node, the trade-off between model complexity and description quality is made in the choice of $r$. We define a simple and intuitive measure of quality of fit to guide this choice in situations where it cannot be made with system knowledge. We define the \textit{flow overlap} of two discrete probability distributions over the same domain as the probability mass in the same elements:
\begin{equation}
    \text{flow overlap}(p, q) := \sum_i \min (p(i) ,q(i)),
    \label{eq:flow_overlap_vector}
\end{equation}
where $p$ and $q$ are discrete probability distributions over a set indexed by $i$. We can extend flow overlap to transition matrices as the (weighted) mean flow overlap of every column. Let $\mathbf{\hat{X}(r)}$ be the rank-$r$ solution. Flow overlap is defined as 
\begin{multline}
 \text{flow overlap}\left( \mathbf{X}, \mathbf{\hat{X}(r)} \; | \; n_i \right)\\
 :=  \frac{1}{\sum_{i'}n_{i'}} \sum_{k,i}  n_i \min \left( X_{ki}, \hat{X}(r)_{ki} \right).
    \label{eq:flow_overlap}    
\end{multline}
This quantity  $\in [0,1]$ measures the fraction of flow through the physical node in $\mathbf{X}$ that is captured by $\mathbf{\hat{X}(r)}$. In our example, flow overlap increases rapidly until rank 3, after which adding state nodes does not capture important new behaviour (Fig.~\ref{fig:methods_schematic}). We pick the optimum number of state nodes as the lowest $r$ such that the flow overlap reaches a threshold.

\paragraph*{Constructing the network.} State nodes can be made independently in parallel for each physical node. In large systems, we can also restrict the creation of state nodes to a subset of important physical nodes. We put the concise network model together by linking state node $\alpha^i$ to $\beta^j$ with weight
\begin{align}
     \hat{P}(\alpha^i \rightarrow \beta^j) &= (\hat{X}^i_\text{out})_{j \alpha^i} (\hat{X}^j_\text{in})_{i \beta^j} \nonumber \\
     & = \hat{P}(\alpha^i \rightarrow j) \hat{P}(i \rightarrow \beta^j),
\end{align}
where the superscript indicates the physical node. Convex NMF tends to create sparse factors \cite{ding2008convex}. Many of the entries of $\mathbf{\hat{X}_\text{in}}$ and $\mathbf{\hat{X}_\text{out}}$ will be close to $0$, meaning that predecessors and successors interact primarily with a subset of state nodes. However, since the NMF is likely to converge before they are exactly $0$, state nodes have dense neighbourhoods with many low-weight edges. We recommend trimming these low-importance edges, for which we implement a simple threshold-based approach (see Methods).

\subsection*{Synthetic experiments}
\begin{figure}
    \centering
    \includegraphics[width=\linewidth]{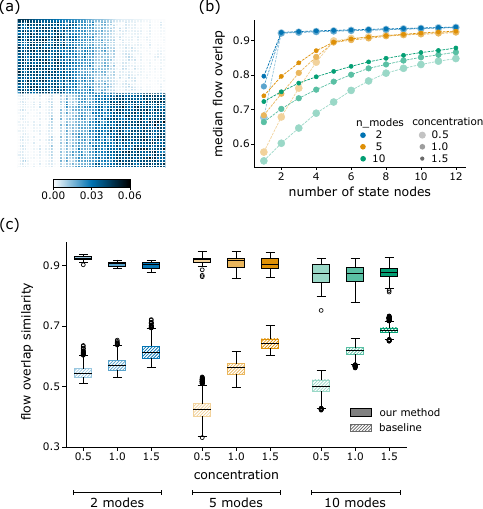}
    \caption{\textbf{Synthetic experiments.} \textbf{(a)} A typical example of $\mathbf{X}$ for 2 modes and $\texttt{concentration}=0.5$. Predecessors are ordered by their participation in the first mode. \textbf{(b)} Median flow overlap as a function of the number of state nodes. \textbf{(c)} Flow overlap similarity between the state nodes created by our method (solid) and the baseline (hatched) with the predecessors closest to the planted modes. We distinguish $\texttt{n\_modes}$ with colours and $\texttt{concentration}$ with transparency.}
    \label{fig:dirichlet_figure}
\end{figure}

We explore the performance of our method on synthetic physical nodes with 50 predecessors and successors each. We sample data for each predecessor from a distribution that is a random convex combination of $\texttt{n\_modes}$ non-overlapping planted modes. The modes combine with weights drawn from a symmetric distribution with spread controlled by $\texttt{concentration} > 0$. The distribution is uniform for $\texttt{concentration}=1$ and higher values decrease the variance. Intuitively, each predecessor participates exclusively in a single mode for $\texttt{concentration}\ll 1$, and equally in all the modes for $\texttt{concentration}\gg 1$. We vary $\texttt{n\_modes} \in \{ 2,5,10\}$ and $\texttt{concentration} \in \{ 0.5, 1.0, 1.5 \}$, and generate 25 physical nodes for each pair (see Methods for details).

When $\texttt{n\_modes}=2,5$, flow overlap (Eq.~\ref{eq:flow_overlap}) increases rapidly until the number of state nodes equals $\texttt{n\_modes}$, reaching a high value that does not improve further with more state nodes (Fig.~\ref{fig:dirichlet_figure}(b)). This sharp increase to an optimum is less clear when there are 10 modes. Nonetheless, flow overlap reaches reasonably high values of more than $0.8$ with 10 state nodes. As expected, solutions with fewer than $\texttt{n\_modes}$ state nodes do better for higher $\texttt{concentration}$ since the predecessors are more similar.

Since the planted modes are extreme behaviours unlikely to be observed in the data, we do not want the state nodes to match them exactly. Instead, we compare the state nodes to the predecessors which are most similar to the modes. We calculate the quality of a solution as the mean flow overlap (Eq.~\ref{eq:flow_overlap_vector}) of the best matching. As a baseline, we randomly generate 50 solutions with the same number of state nodes (see Methods). Not only do we out-perform the baseline for all values of the parameters, but we also achieve objectively high values of similarity (Fig.~\ref{fig:dirichlet_figure}(c)).

\subsection*{Transit flow through airports}
While second-order memory is important in modelling the flow of passengers through airports \cite{rosvall2014memory, salnikov2016using, larock2020hypa}, fixed second-order networks can be impractically large and are prone to overfitting. Here, we investigate whether we can capture key memory effects with interpretable state nodes and if our concise network model offers insights different from a first-order network. We use open source data from the U.S. Bureau of Transport Statistics to construct a dataset of around 4.3 million domestic transits through 435 airports in the United States. We are interested specifically in the role of airports as transit hubs and only consider non-return transits of the form $i \rightarrow j \rightarrow k$ where $i \neq k$ (see Methods).

\begin{figure}[!h]
    \centering
    \includegraphics[width=\linewidth]{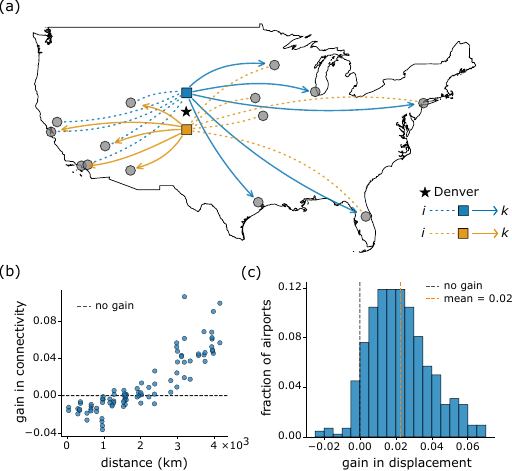}
    \caption{\textbf{State nodes and connectivity in the airport network.} \textbf{(a)} Denver with two state nodes. For either state node, we plot with grey circles the 5 predecessors (resp. successors) with the highest $(\hat{X}_\text{in})_{i\alpha}$ (resp. $(\hat{X}_\text{out})_{k\alpha}$) from amongst the 20 most-observed predecessors (resp. successors) of Denver. The state nodes are denoted as blue and orange squares with edges of the same colour. Dashed (resp. solid) lines are edges to (from) state nodes. The black star marks the location of Denver. \textbf{(b)} Gain in 3-leg connectivity (y-axis) -- $\log_{10}\left( \rho_\text{c} / \rho_\text{fo} \right)$ -- against the distance between $o$ and $d$ (x-axis) for $(o,d)$ pairs of airports ranked 11-20 by number of transits. The dashed black line denotes no gain. \textbf{(c)} The distribution of gain in 3-leg displacement -- $\log_{10}\left( \delta_\text{c} / \delta_\text{fo} \right)$ -- for all origin airports except the 10 largest hubs. The dashed black line denotes no gain. The dashed orange line marks the mean.}
    \label{fig:flights_figure}
\end{figure}

The five largest airports by transit volume -- in Atlanta, Dallas-Fort Worth, Denver, Charlotte, and Chicago -- account for 42\% of all transits, making it crucial to capture potential memory effects in their traffic. However, they have an average of 170 connections each, and a full second-order model is impractical. We explore whether we can build a more concise model by identifying meaningful modes of behaviour (see SI Sec.~\ref{sisubsec:flights_5_largest}). Flow overlap increases rapidly between rank 1 and rank 2 or 3, indicating large-scale patterns that can be modelled with only a few state nodes. For instance, having two state nodes for Denver increases flow overlap from $0.60$ to $0.74$ by capturing an intuitive behaviour -- passengers arriving from the east are likely to continue westward and vice-versa (Fig.~\ref{fig:flights_figure}(a)). We observe similar patterns for the other large national hubs, revealing that passengers use them to travel between distant regions.

The inability of a first-order network to model this behaviour suggests that adding memory effects would alter the analysis of connectivity, which is an important feature of any transport system. We analyse this by constructing (1) a first-order network $G_\text{fo}$ and (2) a concise memory network $G_\text{c}$ with state nodes for the 10 largest airports, which account for 57\% of the transits (see Methods). Using a flow overlap threshold of $0.7$, $G_\text{c}$ has 27 state nodes for these airports and is much smaller than a network with full second-order models for them, which would have 1,467 state nodes.

Modelling a passenger's itinerary as a random walk on the network, we define 3-leg connectivity $\rho_\text{fo} (o,d)$ (resp. $\rho_\text{c} (o,d)$) as the probability that a passenger starting at origin $o$ reaches destination $d$ in $3$ or fewer steps on the first-order (resp. concise) network (see Methods). Comparing $G_\text{c}$ to $G_\text{fo}$, the gain in connectivity is $\log_{10}\left( \rho_\text{c} / \rho_\text{fo} \right)$. For $(o,d)$ pairs in the next 10 largest airports, the gain (1) increases with the geographic distance between $o$ and $d$ (Pearson $r=0.88$, Kendall $\tau=0.74$, both with p-value $<10^{-16}$), and (2) is positive for all pairs of airports more than 2,500 km apart (Fig.~\ref{fig:flights_figure}(b)). These results are robust to the number of legs (SI Fig.~\ref{sifig:flights_connectivity_bias_sweep}).

Covering larger distances on $G_\text{c}$ is not unique to journeys from big airports. Let $\delta_\text{fo}(o)$ (rep. $\delta_\text{c}(o)$) be the expected displacement from origin $o$ after $3$ steps on $G_\text{fo}$ (resp. $G_\text{c}$). The gain in 3-leg displacement -- $\log_{10}\left(\delta_\text{c}/\delta_\text{fo}\right)$ is positive for most origin airports (Fig.~\ref{fig:flights_figure}(c)). The mean gain $=0.02$ is significantly greater than $0$ (one-tailed t-test statistic$=27.2$, p-value$<10^{-16}$). By capturing the role of large national hubs in routing traffic across distant regions, $G_\text{c}$ shows that passengers can travel long distances in a few flights.

\subsection*{Group structure in information flow}
Information flow is an important process in the analysis of social networks, where individuals are modelled as nodes and their interactions as edges. In reality, people interact in many different contexts, and whom you pass information on to likely depends on whom you got it from. Using social networks of co-work and friendship among 71 lawyers at a firm \cite{lazega2001collegial}, we generate synthetic trajectories where information received from a friend (resp. co-worker) is passed on to a friend (resp. co-worker). From these, we construct (1) the first-order network $G_\text{fo}$, (2) a concise network $G_\text{c}$ with flow overlap threshold $=0.9$, having 2 state nodes each for 52 individuals, and (3) the second-order network $G_\text{so}$ (see Methods).

\begin{figure*}[ht]
    \centering
    \includegraphics[width=\linewidth]{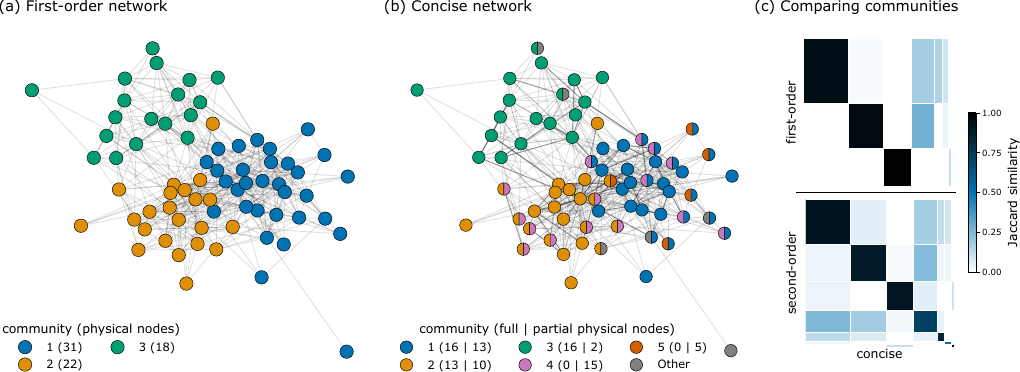}
    \caption{\textbf{Group structure of information flow in the social network.} Communities identified by Infomap in the \textbf{(a)} first-order and \textbf{(b)} concise networks. The circles are physical nodes with colours indicating community membership. The legend includes community sizes, with the two smallest communities with fewer than 5 nodes each in $G_\text{c}$ labelled \textit{Other}. Node positions are determined by the spring layout applied to $G_\text{fo}$. \textbf{(c)} Jaccard similarity of physical nodes in each community of $G_\text{c}$ (x-axis) with those of $G_\text{fo}$ (y-axis, top) and $G_\text{so}$ (y-axis, bottom). The width (resp. height) of each cell is proportional to the size of the community in $G_\text{c}$ (resp. $G_\text{fo}$ and $G_\text{so}$).}
    \label{fig:lazega_figure}
\end{figure*}

Identifying group structures is key to understanding information spread. We use Infomap \cite{mapequation2024software}, a flow-based community detection tool, to find groups of people within which information circulates rapidly before spreading to the rest of the network (see Methods). The three communities of $G_\text{fo}$ (Fig. \ref{fig:lazega_figure}(a)) correlate with work-related metadata. Office location splits the individuals in community 3 from the rest, who are then divided into litigators and corporate lawyers (SI Table~\ref{sitab:lazega_fo_module_metadata_crosstab}).

For higher-order networks, Infomap works at a state node level, allowing communities to overlap at physical nodes~\cite{edler2017mapping}. $G_\text{c}$ has seven communities (Fig.~\ref{fig:lazega_figure}(b)), of which 2 are very small. Communities 1 - 3 are non-overlapping and are the same groups of co-workers identified in $G_\text{fo}$ (Fig.~\ref{fig:lazega_figure}(c), top). Communities 4 and 5 overlap with 1 and 2, and are novel to $G_\text{c}$. They are friendship groups of people who work in the same office, and correspond exactly to communities in the friendship network (SI Sec.~\ref{sisubsec:comms_of_gvo}). This analysis shows that $G_\text{c}$ effectively captures overlapping social circles, revealing distinct yet interconnected groups of friends and co-workers. Despite having more than $7$ times as many nodes, the community structure of $G_\text{so}$ is very similar to that of $G_\text{c}$ (Fig.~\ref{fig:lazega_figure}(c), bottom), showing that we can model the important memory effects with far fewer nodes.

\section*{Conclusion}

We propose a pipeline for constructing concise network models from path data. 
Using non-negative matrix factorisation, we create interpretable state nodes that capture memory effects, identifying large-scale patterns in second-order dynamics.
To prevent overfitting, we incorporate Bayesian regularisation and define a simple measure of model performance to guide the  trade-off between model size and performance.
Our approach produces compact network models that retain important memory effects in empirical data at a fraction of the cost of full second-order models.

This work opens up several promising avenues for future research. The flexibility and scalability of the pipeline make it well-suited for analysing real-world path data across diverse domains. A natural extension is to incorporate the memory effects of higher orders. Similar tools can be explored to identify non-stationary patterns and to construct compact representations of temporal networks.

%TC:ignore
\section*{Methods}
\subsection*{Constructing concise networks}
\subsubsection*{The prior}
We assume that there are sufficient data that $\mathbf{M1}$ is a good estimate of first-order transition rates. This assumption is not essential to our framework. For instance, in cases where data are very sparse, a uniform prior or one conserving node degree might be more suitable. However, data that are insufficient to estimate first-order transition rates are unlikely to be usable for higher-order models.
\paragraph*{Leave-one-out cross-validation to pick $\mu$.}
The choice of the prior strength is critical, as high values of $\mu$ wash out memory effects while low values fail to remove the noise from $\mathbf{M2}$. In cases where system knowledge cannot inform the choice, we use leave-one-out cross-validation \cite{zhai2004}. From Eq.~\ref{eq:Xki}, the likelihood of observing a $i \rightarrow j \rightarrow k$ in a model trained on all the others is
\begin{equation}
    \frac{A_{ki} -1 + \mu \bm{M1}_k}{n_i -1 + \mu}.
\end{equation}
We pick the $\mu$ that maximises the log-likelihood of every observed transition in a model trained on all the others, i.e.
\begin{equation}
    \mu^* = \argmax_\mu \sum_{k,i} A_{ki} \log \left(  \frac{A_{ki} -1 + \mu \bm{M1}_k}{n_i -1 + \mu} \right).
\end{equation}
Our choice of leave-one-out cross-validation is motivated by its closed-form objective function and ease of optimisation. Using other methods such as $k$-fold cross-validation to estimate the parameter(s) of the prior does not affect the rest of our method.

\subsubsection*{Convex NMF to construct state nodes}
NMF is a tool to create low-rank approximations with interpretable factors. The goal of NMF is to obtain factorisations of the form $\mathbf{X} \approx \mathbf{\hat{X}} = \mathbf{FG}^\top$, where $\mathbf{F},\mathbf{G} \geq 0$. Convex NMF constrains the space of solutions by requiring the columns of $\mathbf{F}$ to lie in the column space of $\mathbf{X}$. Thus, we search for a factorisation $\mathbf{\hat{X}} = \mathbf{XWG}^\top$, where $\mathbf{W},\mathbf{G}^\top \geq 0$. We use the multiplicative updates from Ref.~\cite{ding2008convex} to optimise the loss function
\begin{equation}
    ||\mathbf{\hat{X}} - \mathbf{X}||^2 = \sum_{k,i} \left( \hat{X}_{ki} - X_{ki} \right)^2.
    \label{eq:nmf_loss}
\end{equation}
In each iteration, we update
\begin{align*}
    \mathbf{G} &\leftarrow \mathbf{G} \odot \sqrt{ \left(\mathbf{X}^\top \mathbf{X} \mathbf{W}\right) \oslash \left(\mathbf{G} \mathbf{W}^\top \mathbf{X}^\top \mathbf{X} \mathbf{W} \right)} \\
    \mathbf{W} &\leftarrow \mathbf{W} \odot \sqrt{ \left( \mathbf{X}^\top \mathbf{X} \mathbf{G} \right) \oslash \left( \mathbf{X}^\top \mathbf{X} \mathbf{W} \mathbf{G}^\top \mathbf{G} \right) },
\end{align*}
where $\odot$ and $\oslash$ are element-wise multiplication and division respectively until the solution converges to the local optimum (see Ref.~\cite{ding2008convex} for proofs of correctness and convergence). The solution is not unique. For instance, we can write $\mathbf{XWG}^\top = \mathbf{X}(\mathbf{WA}^{-1})(\mathbf{GA}^\top)^\top$ for alternative solutions with the same loss. Requiring that the matrix factors are transition matrices eliminates this degeneracy. We set
\begin{align}
    \mathbf{\hat{X}_\text{out}} &= \mathbf{XWD_W}^{-1}\\
    \mathbf{\hat{X}_\text{in}} &= \mathbf{GD_W},
\end{align}
where $\mathbf{D_W}$ is the diagonal matrix of the column sums of $\mathbf{W}$. The columns of $\mathbf{\hat{X}_\text{out}}$ are convex combinations of the columns of $\mathbf{X}$, making $\mathbf{\hat{X}_\text{out}}$ column-stochastic. The row-sums of $\mathbf{\hat{X}_\text{in}}$ will be close to $1$ and we normalise them. For each $r$, we pick the solution with the lowest loss amongst $\texttt{n\_candidates}$ candidates with different initialisations.

\paragraph*{Initialisation.} We use the equivalence of Convex NMF to soft k-means clustering (SI Sec.~\ref{sisubsec:convexnmf_kmeans}) to pick good initial values. We initialise $\mathbf{G}$ to a smoothed version of the k-means clustering of the columns of $\mathbf{X}$ into $r$ clusters,
\begin{equation}
    G_{i\alpha} = 
        \begin{cases}
            1, & \text{if column } i \in \text{ cluster }\alpha \\
            0.2, & \text{otherwise}.
        \end{cases}
\end{equation}
We initialise $\mathbf{W}$ to a row-normalised version of $\mathbf{G}$.

\paragraph*{Convergence.} The loss is guaranteed to be non-increasing under the multiplicative updates\cite{ding2008convex}. In the analyses in this work, we declare convergence when the relative decrease in loss over 10 iterations is less than $10^{-4}$.

\subsubsection*{Trimming the neighbourhood of state nodes}
We implement a simple method to trim out low-importance edges from the neighbourhood of state nodes, improving the interpretability and sparsity of the network. For a physical node $j$ with rank $r$, the importance of state node $\alpha$ to predecessor $i$ is $\left( \hat{X}_\text{in} \right)_{i\alpha}$ -- the probability that a trajectory from $i$ passes through $\alpha$. We retain edge $i \rightarrow \alpha$ if
\begin{equation}
    \left(\hat{X}_\text{in}\right)_{i\alpha} \geq \frac{1}{r} \times \texttt{multiplier},
\end{equation}
where $\texttt{multiplier}$ is a parameter that controls the strictness of the trimming such that smaller values retain more edges. Similarly, we keep $\alpha \rightarrow k$ if
\begin{equation}
    \frac{ \left(\hat{X}_\text{out}\right)_{k\alpha} } { \sum_\beta \left( \hat{X}_\text{out} \right)_{k\beta} } \geq \frac{1}{r} \times \texttt{multiplier}.
\end{equation}

\subsection*{Synthetic examples}
\subsubsection*{Synthetic data}
We construct physical nodes with 50 predecessors and successors each by generating synthetic flow using $\texttt{n\_modes}$ modes of behaviour. The modes are uniform distributions over disjoint equal-sized subsets of the successors. For instance, when $\texttt{n\_modes}=5$, we partition the successors into 5 groups of 10 each, and define each mode as the uniform distribution over the successors in the corresponding group. For each predecessor, we create a `true' distribution as a convex combination of the modes with weights drawn from a symmetric Dirichlet distribution with concentration parameter $\texttt{concentration}$. Increasing $\texttt{concentration}$ decreases variance, making predecessors more similar on average. We generate the data for each predecessor by sampling its `true' distribution $1000$ times. We vary $\texttt{n\_modes} \in \{2,5,10\}$ and $\texttt{concentration} \in \{ 0.5, 1.0, 1.5 \}$, constructing 25 instances for each combination.

\subsubsection*{Quality of state nodes}
Predecessors are likely to be a mixture of modes. Particularly for larger values of $\texttt{concentration}$, the planted modes are not realistic behaviour given the data. Therefore, it is undesirable for the state nodes to match them exactly. We evaluate the quality of the state nodes by comparing them to the set of predecessors that are most similar to the planted modes.

For each mode, we identify the closest predecessor by finding the column of $\mathbf{X}$ with the highest flow overlap. Using the same similarity measure, we find the best matching of the state nodes to these predecessors and calculate the mean flow overlap similarity of the matched pairs.

\paragraph*{The baseline.} We compare the quality of our solution with randomly generated ones with the same number of state nodes. Each state node in the baseline is a convex combination of the columns of $\mathbf{X}$ with weights distributed uniformly at random. For each physical node, we generate 50 instances of the baseline.

\subsection*{Transit flow through airports}
\subsubsection*{Data description}
Our dataset is a $10\%$ sample of domestic flight itineraries in the United States in the first quarter of 2023\footnote{Airline Origin and Destination Survey, Bureau of Transportation Statistics, accessed on 13 June 2024. \url{https://www.transtats.bts.gov/DataIndex.asp}}. We discard all itineraries with just one leg and parse the rest into transits through each airport. Since we are interested in the role of airports as transit hubs, we remove \textit{return} transits of the form $i \rightarrow j \rightarrow i$. This leaves 4,340,809 transits between 435 airports. Of these, 379 have transits through them, while the rest only serve as sources or destinations. We plot the distribution of transit volume (SI Fig.~\ref{sifig:flights_hist_ntransits}) and the location of the airports (SI Fig.~\ref{sifig:flights_map_all_transit}) in the SI.

\subsubsection*{Constructing the networks}
We create the first-order network $G_\text{fo}$ by setting rank $=1$ for each physical node. For the concise memory network $G_\text{c}$, we create state nodes for the 10 largest transit hubs (SI Table~\ref{sitab:flights_top10_airports}) with flow overlap threshold $0.7$. $G_\text{c}$ has 2 state nodes each for Atlanta, Dallas-Fort Worth, Denver, Charlotte, Chicago, Seattle, and Houston, 3 for Minneapolis - St Paul, and 5 each for Phoenix and Las Vegas.

\paragraph*{Backboning.} We remove low-importance edges in two stages. First, we trim the neighbourhoods of the state nodes in $G_\text{c}$ with $\texttt{multiplier}=0.05$. Next, we use the Disparity Filter \cite{serrano2009extracting} to backbone both networks\footnote{We use the implementation by Michele Coscia available at \url{https://www.michelecoscia.com/?page_id=287}.} with a DF score threshold of $0.01$ (see SI Sec.~\ref{sisubsec:flights_constructing_networks}). We ensure that both networks remain weakly connected. After backboning, $G_\text{fo}$ has 435 nodes and 9,249 edges, and $G_\text{c}$ has 452 nodes and 12,429 edges.

\subsubsection*{Connectivity analysis}
We model an itinerary as a discrete time random walk on the network. Let $\mathbf{T_\text{fo}}$ be the row-stochastic adjacency matrix of $G_\text{fo}$\footnote{We give the three airports with no out-edges self-loops with weight $=1$.}.
\paragraph*{3-leg connectivity.} The probability that a passenger at origin $o$ reaches destination $d$ in $3$ or fewer steps on $G_\text{fo}$ is given by
\begin{equation}
    \rho_\text{fo}(o,d) := \left( \left(\mathbf{T^d_\text{fo}} \right)^3 \right)_{od},
\end{equation}
where $\mathbf{T^d_\text{fo}}$ is $\mathbf{T_\text{fo}}$ modified to make $d$ an absorbing state. Specifically, $\left(T^d_\text{fo}\right)_{id}=0$ $\forall i \neq d$ and $\left(T^d_\text{fo}\right)_{dd}=1$. We define $\rho_\text{c}(o,d)$ similarly for cases where $o$ and $d$ have only one state each. In the Results, we investigate the gain in 3-leg connectivity between $(o,d)$ pairs from the airports ranked 11 - 20 by transit volume (SI Table~\ref{SItab:flights_next10_airports}).
\paragraph*{3-leg displacement.} The expected 3-leg displacement from $o$ for $G_\text{fo}$ is
\begin{equation}
    \delta_\text{fo} (o) := \sum_d \left( \left( \mathbf{T_\text{fo}} \right) ^3\right)_{od} \times \text{distance}(o,d).
\end{equation}
We define $\delta_\text{c}$ similarly. In the Results, we investigate the gain in 3-leg displacement for all origins $o$ except the 10 largest airport. In both analyses, we use Haversine distance to quantify geographic separation.

\subsection*{Group structure in information spread}
\subsubsection*{Data description}
The Lazega law firm data \cite{lazega2001collegial} contains social networks of three relationships -- co-work, friendship, and advice -- between 71 lawyers at a corporate law firm. We use metadata on the practice (litigation or corporate) and office location (Boston, Hartford, or Providence) of the individuals (SI Table~\ref{sitab:lazega_metadata_crosstab}).

\subsubsection*{Synthetic trajectories}
We start with two separate social networks $G_\text{w}$ of co-work and $G_\text{f}$ of friendship. For simplicity, we make them undirected by discarding non-reciprocated edges. $G_\text{w}$ and $G_\text{f}$ contain 378 and 176 edges respectively, of which 74 are shared. We generate trigrams $i \rightarrow j \rightarrow k$ through every node $j$ using a second-order Markov process to model information spreading separately along co-work and friendship links. If $i$ is only a co-worker (resp.\ friend) of $j$, $k$ is chosen uniformly at random from the set of co-workers (resp.\ friends). If $i$ is both, $k$ is picked from the disjoint union of friends and co-workers. For each $j$, we generate 1000 trigrams from each $i$.

\subsubsection*{Constructing the networks}
We create $G_\text{fo}$ by setting rank $=1$ for each physical node. For $G_\text{c}$, we pick a flow overlap threshold of $0.9$ and create two state nodes each for 52 nodes. The full second-order network $G_\text{so}$ has $|i|$ state nodes for each physical node, with $\mathbf{\hat{X}_\text{out}} = \mathbf{X}$ and $\mathbf{\hat{X}_\text{out}} = \mathbf{I}$, the identity matrix. We trim the neighbourhoods of the state nodes in both higher-order networks with $\texttt{multiplier}=0.05$. $G_\text{fo}$ has 71 nodes and 960 edges, $G_\text{c}$ has 123 nodes and 1,293 edges, and $G_\text{so}$ has 960 nodes and 12,384 edges.

\subsubsection*{Community structure}
Community detection is a long-studied task in network science with a plethora of approaches \cite{fortunato2010community}. We use Infomap \cite{mapequation2024software}, a method designed for networks of flow, and set its Markov time parameter to $0.9$ (see SI Sec.~\ref{sisubsec:lazega_community_detection}).

Infomap works at a state node level, allowing physical nodes to belong to multiple communities. Viewing a community as a set of physical nodes, we compare communities in different networks (Fig.~\ref{fig:lazega_figure}(c)) using Jaccard similarity. For sets $A$ and $B$,
\begin{equation}
    \text{Jaccard similarity} (A, B) := \frac{|A \cap B|}{|A \cup B|}.
\end{equation}

\section*{Acknowledgements}
We thank Jelena Smiljani\'c, Nicola Pedreschi, Rub\'en Bernardo Madrid, and Timothy LaRock for helpful discussions. We acknowledge using open-source software by Michele Coscia, airline data maintained by the U.S. Bureau of Transportation Statistics, and the Lazega network structure from the repository of Manlio De Domenico.

\textbf{Code availability:}
The pipeline is implemented in Python and available at \url{https://github.com/rohit-sahasrabuddhe/concise-networks}. It depends on standard open-source libraries including NumPy~\cite{harris2020array}, pandas~\cite{mckinney-proc-scipy-2010}, NetworkX~\cite{SciPyProceedings_11}, and SciPy~\cite{virtanen2020scipy}.

\textbf{Funding:}
RS is funded by the Mathematical Institute at the University of Oxford. 
RL acknowledges support from the EPSRC grants EP/V013068/1, EP/V03474X/1 and EP/Y028872/1.
MR was supported by the Swedish Research Council under grant 2023-03705.

\textbf{Author contributions:} 
RS: conceptualisation, code implementation, experiment design, experiment execution, interpretation of results, creating figures and plots, writing.
RL: conceptualisation, experiment design, interpretation of results, writing, supervision.
MR: conceptualisation, experiment design, interpretation of results, writing, supervision.
All authors read and approved the final manuscript.

\textbf{Competing interests: }The authors declare no competing interests.

\textbf{Corresponding author: }Rohit Sahasrabuddhe (rohit.sahasrabuddhe@maths.ox.ac.uk)
\bibliographystyle{apsrev4-1}
\bibliography{references}
\newpage
\newpage

\counterwithin{figure}{section}
%\counterwithin{equation}{section}
%\counterwithin{table}{section}
\renewcommand\figurename{Supplementary Figure}
\renewcommand\thefigure{\arabic{figure}}
\renewcommand\tablename{Supplementary Table}
\renewcommand\thetable{\arabic{table}}
{\huge Supplementary Information}
\section{Constructing state nodes}
\subsection{Convex NMF and soft k-means clustering}\label{sisubsec:convexnmf_kmeans}
K-means clustering can be written as a matrix factorisation problem, where we want to approximate the data matrix $\mathbf{X}$ as
\begin{equation}
    \mathbf{X} \approx \mathbf{\hat{X}} = \mathbf{F}\mathbf{G}^\top,
\end{equation}
where $\mathbf{F}$ has the cluster centroids and $\mathbf{G}$ the memberships. Convex NMF is analogous to a soft k-means clustering, with cluster centroids in the convex hull of the data and fuzzy cluster membership. Ref.~\cite{ding2005equivalence} has a general discussion of the relationship between clustering and NMF.

\section{Flight transits network}
\begin{figure}[ht]
    \centering
    \includegraphics[width=\linewidth]{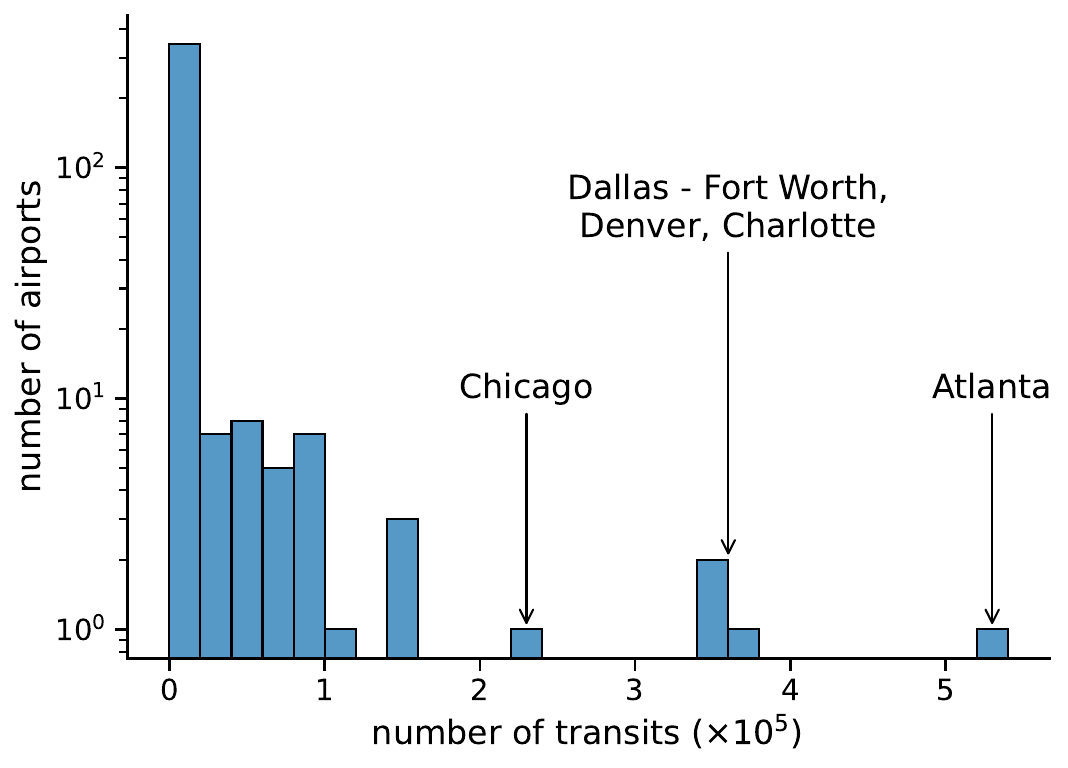}
    \caption{\textbf{Distribution of transit volume.}}
    \label{sifig:flights_hist_ntransits}
\end{figure}
\begin{figure}[ht]
    \centering
    \includegraphics[width=\linewidth]{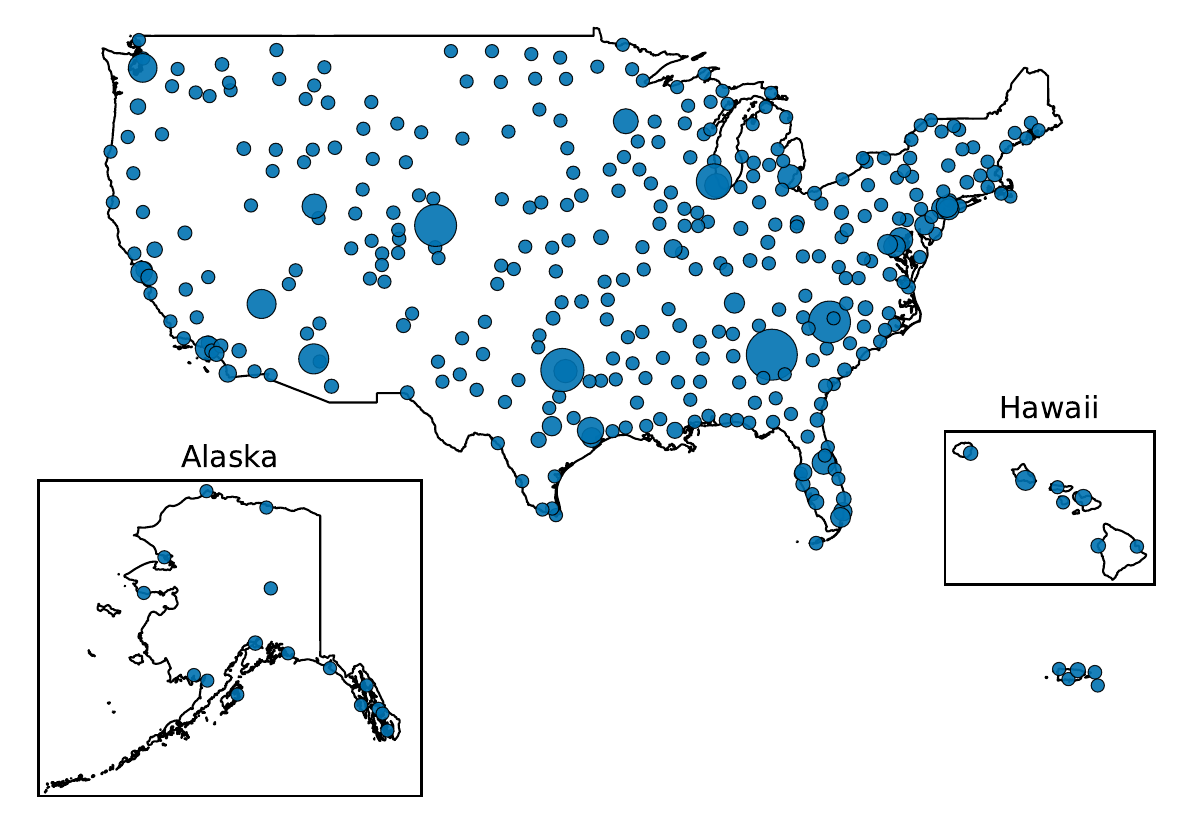}
    \caption{\textbf{Map of airports.} A map of all the airports with at least one transit through them. The size of the marker indicates the number of transits. For a compact map, we place Alaska and Hawaii as insets and do not plot the Virgin Islands, Guam, American Samoa, and Commonwealth of the Northern Mariana Islands.}
    \label{sifig:flights_map_all_transit}
\end{figure}
\subsection{State nodes of the 5 largest hubs}\label{sisubsec:flights_5_largest}
\begin{figure}[ht]
    \centering
    \includegraphics[width=\linewidth]{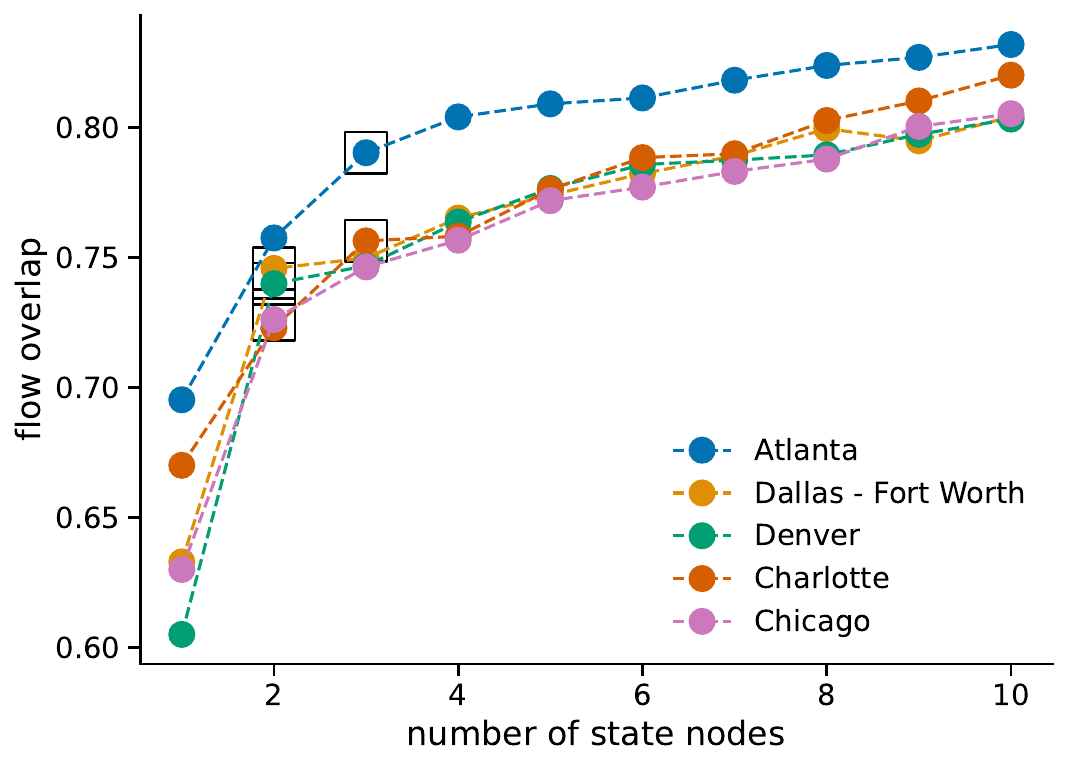}
    \caption{\textbf{Creating state nodes for large airports.} Flow overlap (y-axis) as a function of the number of state nodes (x-axis) for the five largest airports.}
    \label{sifig:flights_top5_flowoverlap}
\end{figure}
\begin{figure}
    \centering
    \includegraphics[width=\linewidth]{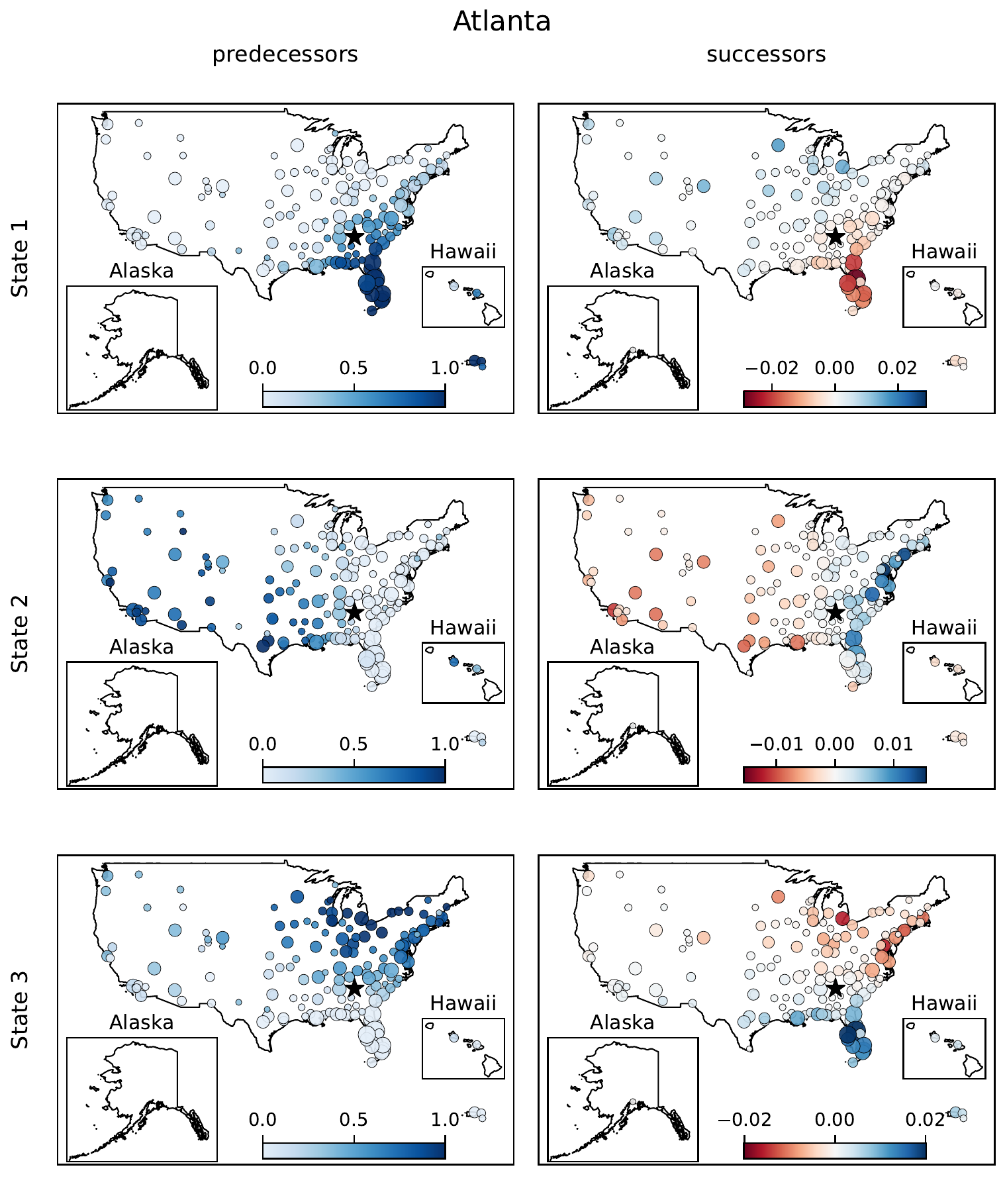}
    \caption{\textbf{State nodes of Atlanta} Predecessors and successors are sized by their traffic with the physical node and coloured by their importance to the state node. We explain this in detail in the text.}
    \label{sifig:flights_ATL_map}
\end{figure}
\begin{figure}
    \centering
    \includegraphics[width=\linewidth]{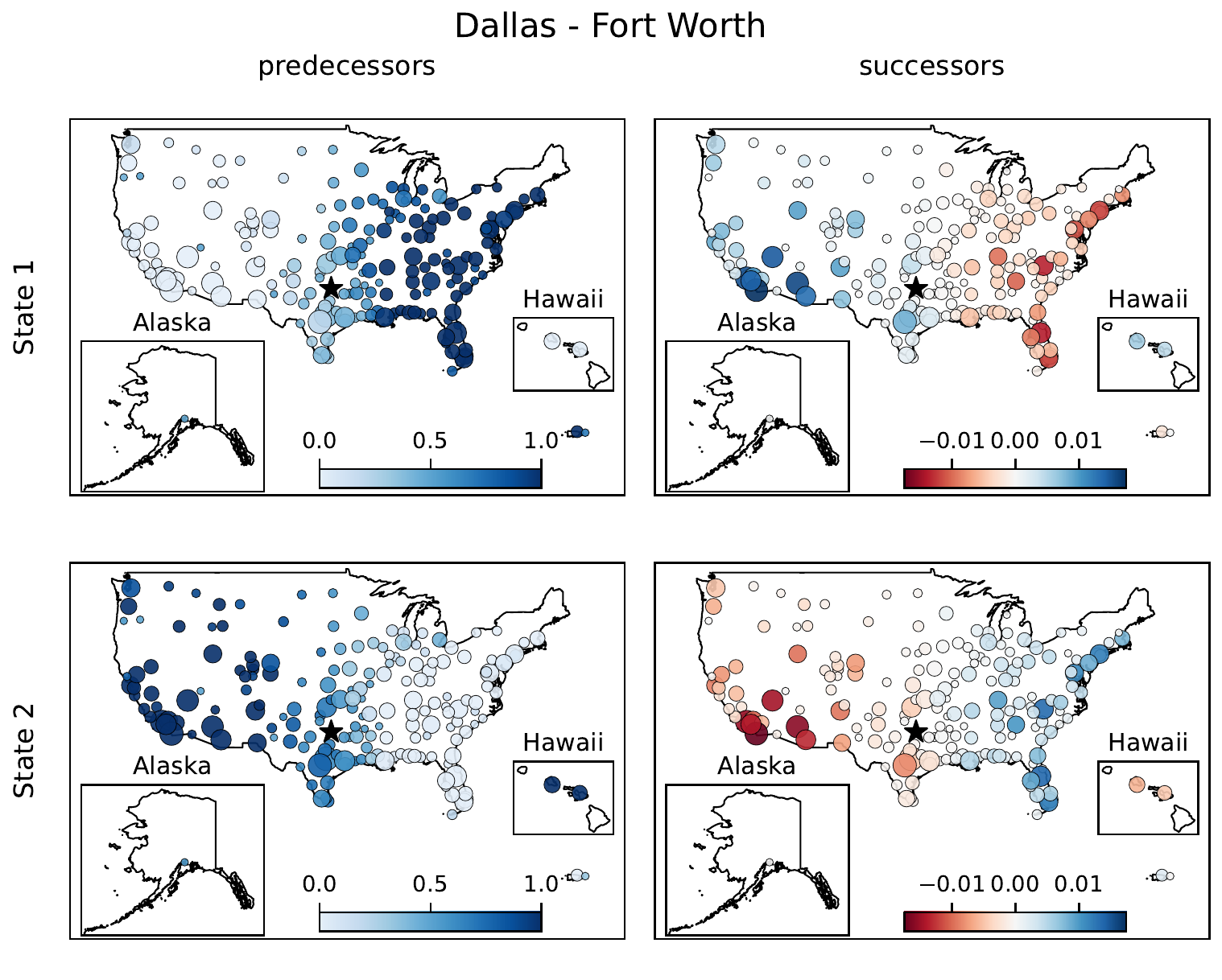}
    \caption{\textbf{State nodes of Dallas - Fort Worth} Predecessors and successors are sized by their traffic with the physical node and coloured by their importance to the state node. We explain this in detail in the text.}
    \label{sifig:flights_DFW_map}
\end{figure}
\begin{figure}
    \centering
    \includegraphics[width=\linewidth]{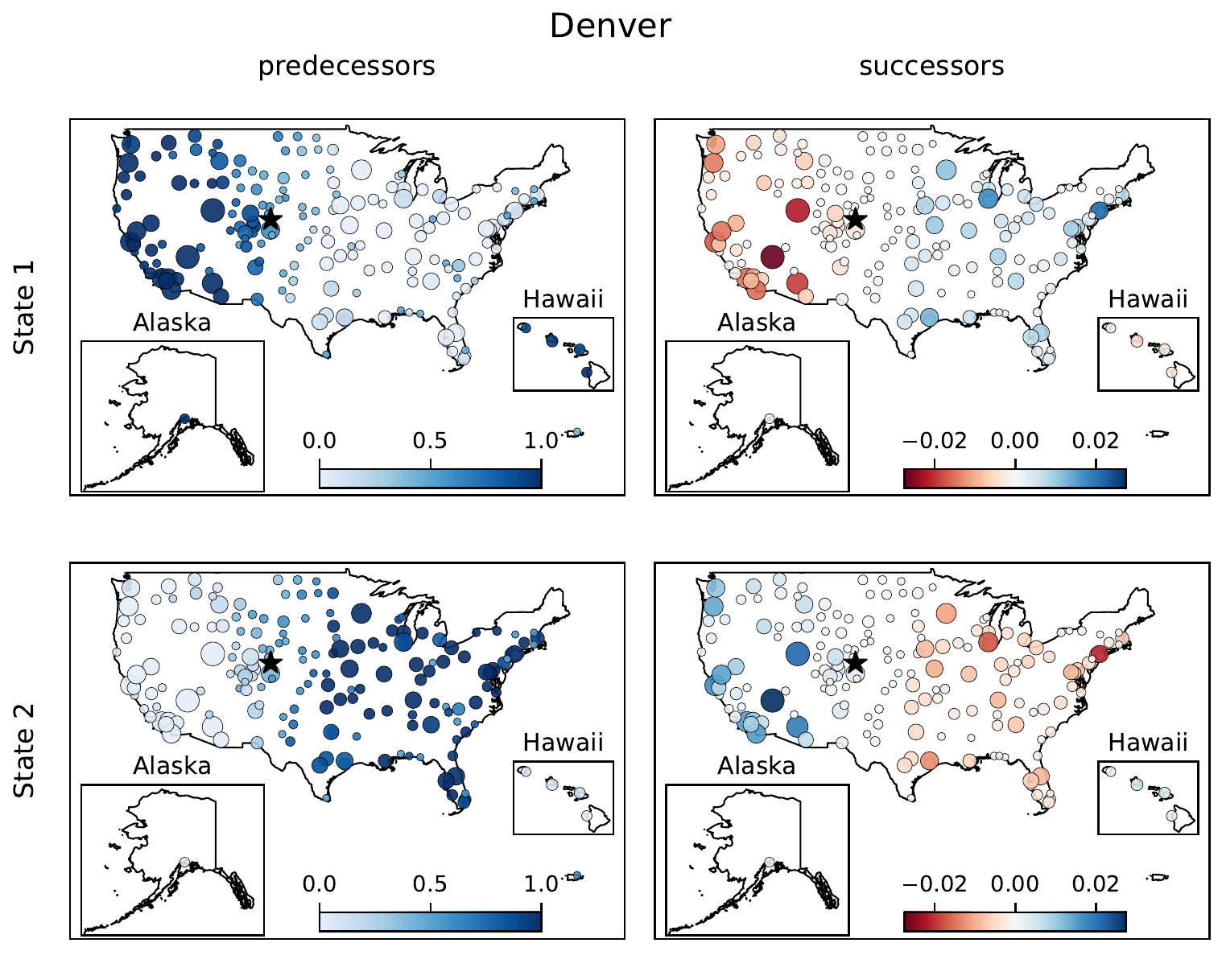}
    \caption{\textbf{State nodes of Denver} Predecessors and successors are sized by their traffic with the physical node and coloured by their importance to the state node. We explain this in detail in the text.}
    \label{sifig:flights_DEN_map}
\end{figure}
\begin{figure}
    \centering
    \includegraphics[width=\linewidth]{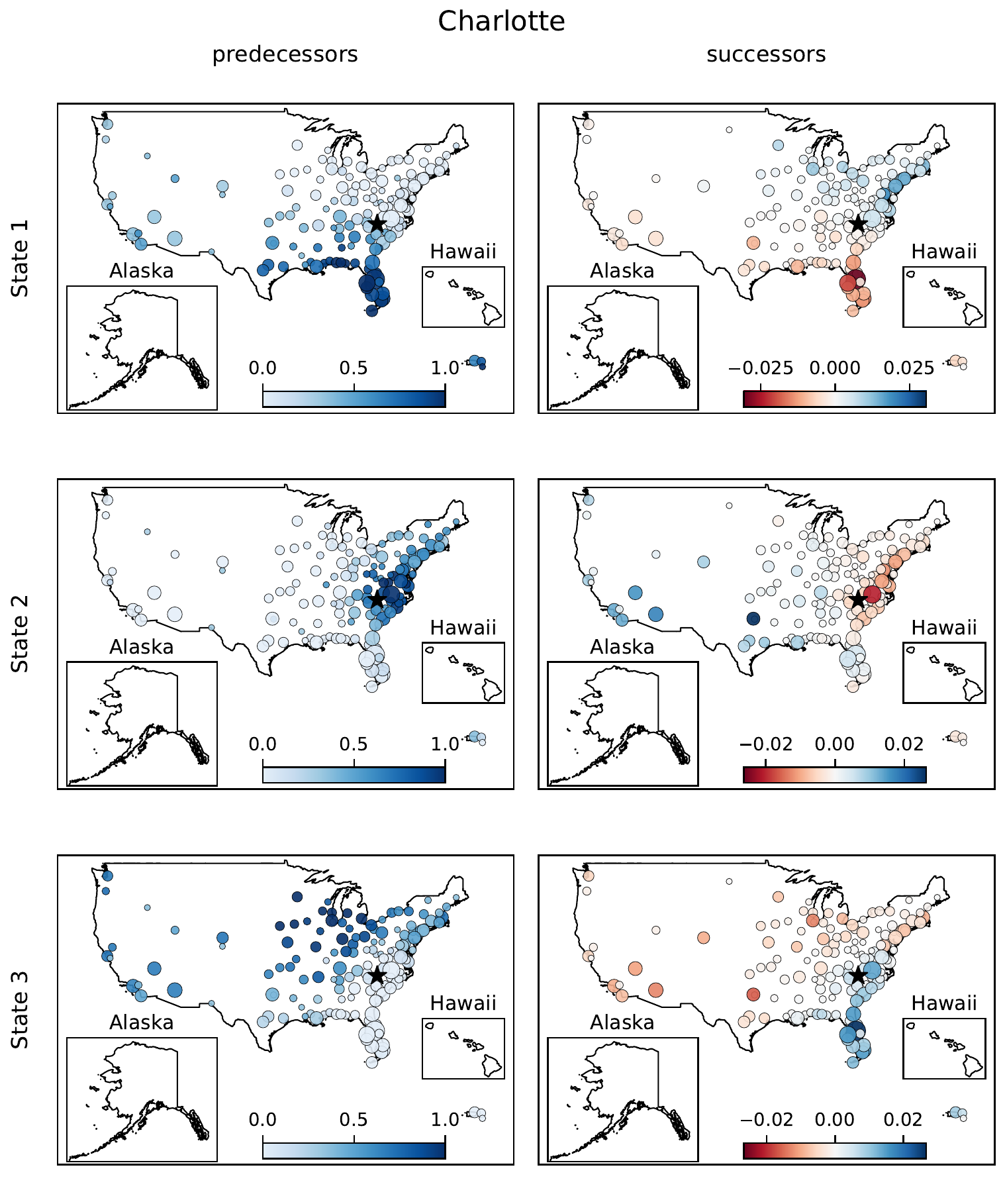}
    \caption{\textbf{State nodes of Charlotte} Predecessors and successors are sized by their traffic with the physical node and coloured by their importance to the state node. We explain this in detail in the text.}
    \label{sifig:flights_CLT_map}
\end{figure}
\begin{figure}
    \centering
    \includegraphics[width=\linewidth]{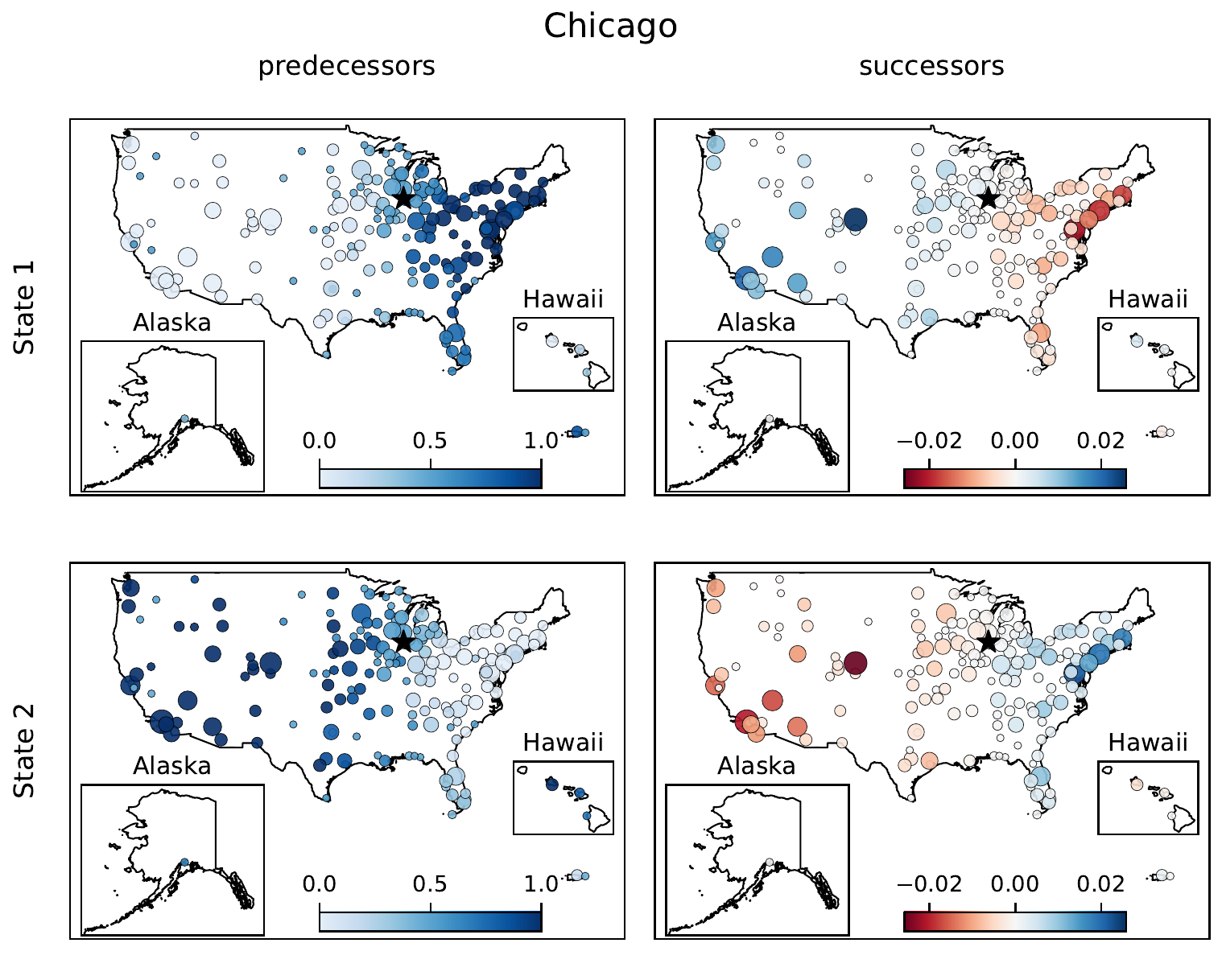}
    \caption{\textbf{State nodes of Chicago} Predecessors and successors are sized by their traffic with the physical node and coloured by their importance to the state node. We explain this in detail in the text.}
    \label{sifig:flights_ORD_map}
\end{figure}

We create models for the five largest airports with 1 - 10 state nodes. We see (1) a steep increase in flow overlap between ranks 1 and 2, and (2) quite high values of flow overlap by rank 3 (SI Fig.~\ref{sifig:flights_top5_flowoverlap}), indicating that important memory effects can be captured by a few large-scale patterns. To illustrate these patterns, we visualise solutions with manually picked number of state nodes --  rank 2 for Dallas, Denver, and Chicago and rank 3 for Atlanta and Charlotte (Figs \ref{sifig:flights_ATL_map} - \ref{sifig:flights_ORD_map}).
\paragraph{}Each row depicts the locations of a state node's predecessors (left) and successors (right). Its marker size indicates the flow from (resp. to) airport to (resp. from) the hub. The colour of $i$ indicates $(\hat{X}_\text{in})_{i\alpha}$. The colour of $k$ shows how over-represented it is in the out-distribution of $\alpha$ compared to the other state nodes:
\begin{equation}
    (\hat{X}_\text{out})_{k\alpha} - \frac{\sum_{\alpha '} (\hat{X}_\text{out})_{k\alpha '}}{r},
\end{equation}
where $r$ is the number of state nodes.
The maps show clear geographic patterns that capture intuitive behaviour -- passengers use large national hubs to travel between distant regions. Travellers staying within a region are probably served by smaller regional hubs or direct flights.

\subsection{Constructing the networks}\label{sisubsec:flights_constructing_networks}
\begin{table*}
    \centering
    \begin{tabular}{c|c|c|c|c}
        IATA code & Airport                                  & \# transits & \# state nodes & flow overlap\\ \hline
        ATL       & Hartsfield-Jackson Atlanta & 529,581    & 2 & 0.76\\
        DFW       & Dallas/Fort Worth          & 369,169    & 2 & 0.75\\
        DEN       & Denver                     & 345,731    & 2 & 0.74\\
        CLT       & Charlotte Douglas          & 343,534    & 2 & 0.72\\
        ORD       & Chicago O'Hare             & 235,590    & 2 & 0.73\\
        PHX       & Phoenix Sky Harbor         & 159,261    & 5 & 0.70\\
        LAS       & Harry Reid                 & 145,290    & 5 & 0.70\\
        SEA       & Seattle/Tacoma             & 140,914    & 2 & 0.72\\
        IAH       & George Bush Intercontinental     & 116,158    & 2 & 0.72\\
        MSP       & Minneapolis-St Paul        & 98,626     & 3 & 0.73
    \end{tabular}
    \caption{\textbf{Largest transit hubs.} The 10 largest airports with number of state nodes and flow overlap in $G_\text{c}$.}
\label{sitab:flights_top10_airports}
\end{table*}
\begin{figure*}
    \centering
    \includegraphics[width=0.8\linewidth]{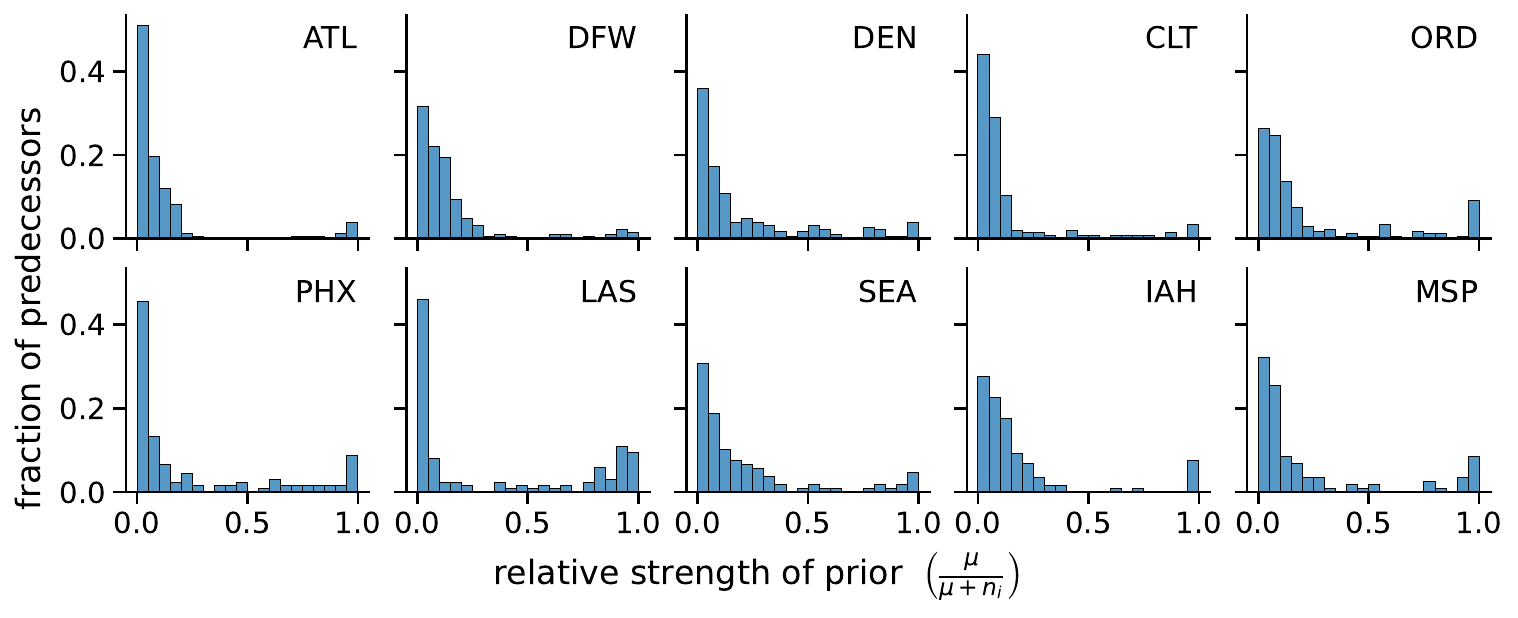}
    \caption{\textbf{Distribution of relative strength of prior}}
    \label{sifig:flights_top10_prior}
\end{figure*}
We construct $G_\text{fo}$ with 1 state node for each airport, and $G_\text{c}$ with state nodes for the 10 largest airports (flow overlap threshold $=0.7$, Table\ref{sitab:flights_top10_airports}). Peaks near 1 in the distributions of the relative strength of the prior (SI Fig.~\ref{sifig:flights_top10_prior}) show that $\mathbf{M2}$ overfits to some predecessors. 

\paragraph{Trimming edges} At this stage, $G_\text{fo}$ has 435 nodes and 10,222 edges, and $G_\text{c}$ has 452 nodes and 15,233 edges. We first trim the neighbourhoods of the state nodes in $G_\text{c}$ using $\texttt{multiplier}=0.05$, which removes 1,309 edges. Next, we apply the Disparity Filter \cite{serrano2009extracting} (implementation by Michele Coscia\footnote{\url{https://www.michelecoscia.com/?page_id=287}}) to remove low-importance edges from both networks.

\begin{figure}
    \centering
    \includegraphics[width=\linewidth]{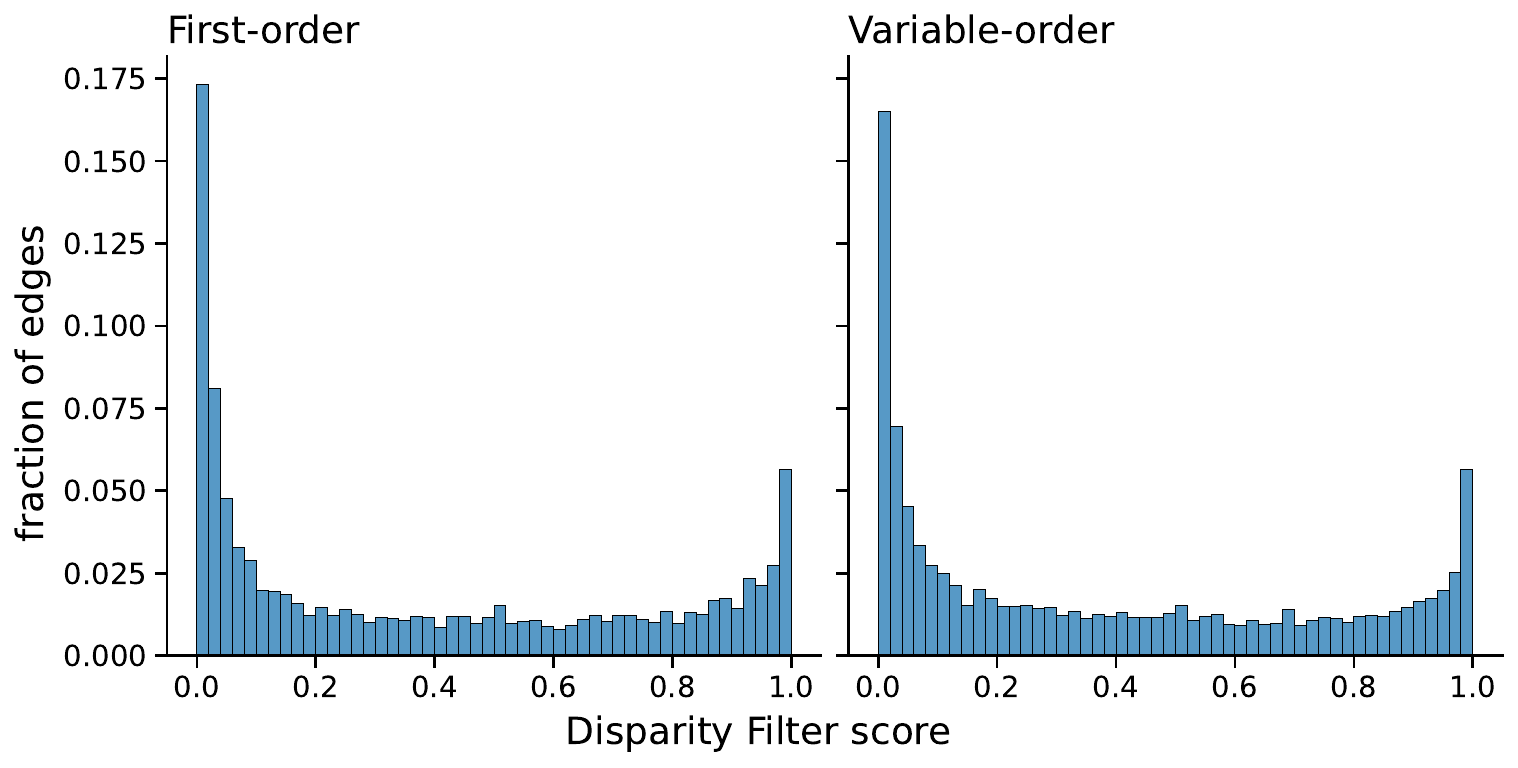}
    \caption{\textbf{Distribution of DF score} Distribution of the Disparity Filter score of the edges in the first-order (left) and concise (right) networks.}
    \label{sifig:flights_dist_dfscore}
\end{figure}

\begin{figure}
    \centering
    \includegraphics[width=\linewidth]{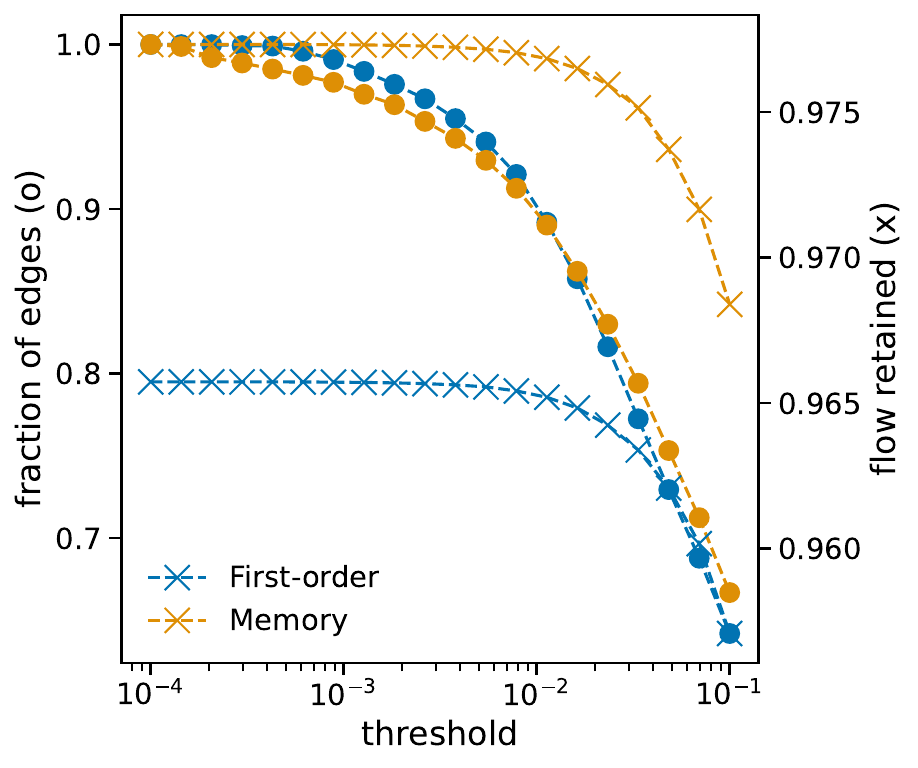}
    \caption{\textbf{Sweeping DF score threshold} The fraction of edges (left x-axis, circle markers) and flow (right x-axis, cross markers) retained as a function of threshold for the first-order (blue) and concise (orange) networks.}
    \label{sifig:flights_df_threshold_sweep}
\end{figure}

\paragraph{Backboning} The DF assigns a score to an edge by comparing its weight to a degree-preserving null model that randomises the weight distribution of the edges of a node. We see a peak near $0$ in the distribution of DF score for both networks \ref{sifig:flights_dist_dfscore}.  Let us define the \textit{flow} along edge $i \rightarrow j$ as $f_{ij} = \text{PageRank}(i) \times w_{ij}$, where $w_{ij}$ is the weight. This can be viewed as the amount of probability mass moving along an edge at stationary state\footnote{Since the networks are not strongly connected, we proxy the stationary state distribution with PageRank ($\alpha=0.85$).}. Thus, $\sum_{i,j} f_{ij}=1$. In SI Fig.~\ref{sifig:flights_df_threshold_sweep}, we plot the fraction of edges and flow retained for DF score threshold $\in [10^{-4}, 10^{-1}]$. With a threshold of $0.01$, $G_\text{fo}$ has 435 nodes and 9,249 edges, and $G_\text{c}$ has 452 nodes and 12,429 edges. We ensure that both networks remain weakly connected.

\begin{table}
\centering
\begin{tabular}{c|c|c}
IATA code & Airport  & \# transits \\ \hline
SLC   & Salt Lake City  & 93,889  \\
LAX  & Los Angeles & 93,849\\
BWI  & Baltimore/Washington  & 90,492 \\
DTW  & Detroit Metro Wayne County  & 88,438\\
MDW  & Chicago Midway   & 87,508\\
DAL  & Dallas Love Field  & 80,362 \\
MCO  & Orlando   & 78,957 \\
LGA  & LaGuardia   & 67,595\\
DCA  & Ronald Reagan Washington & 66,595\\
SFO  & San Francisco   & 64,417
\end{tabular}
\caption{\textbf{Airports ranked 11-20 by transit volume}}
\label{SItab:flights_next10_airports}
\end{table}

\begin{figure}
    \centering
    \includegraphics[width=\linewidth]{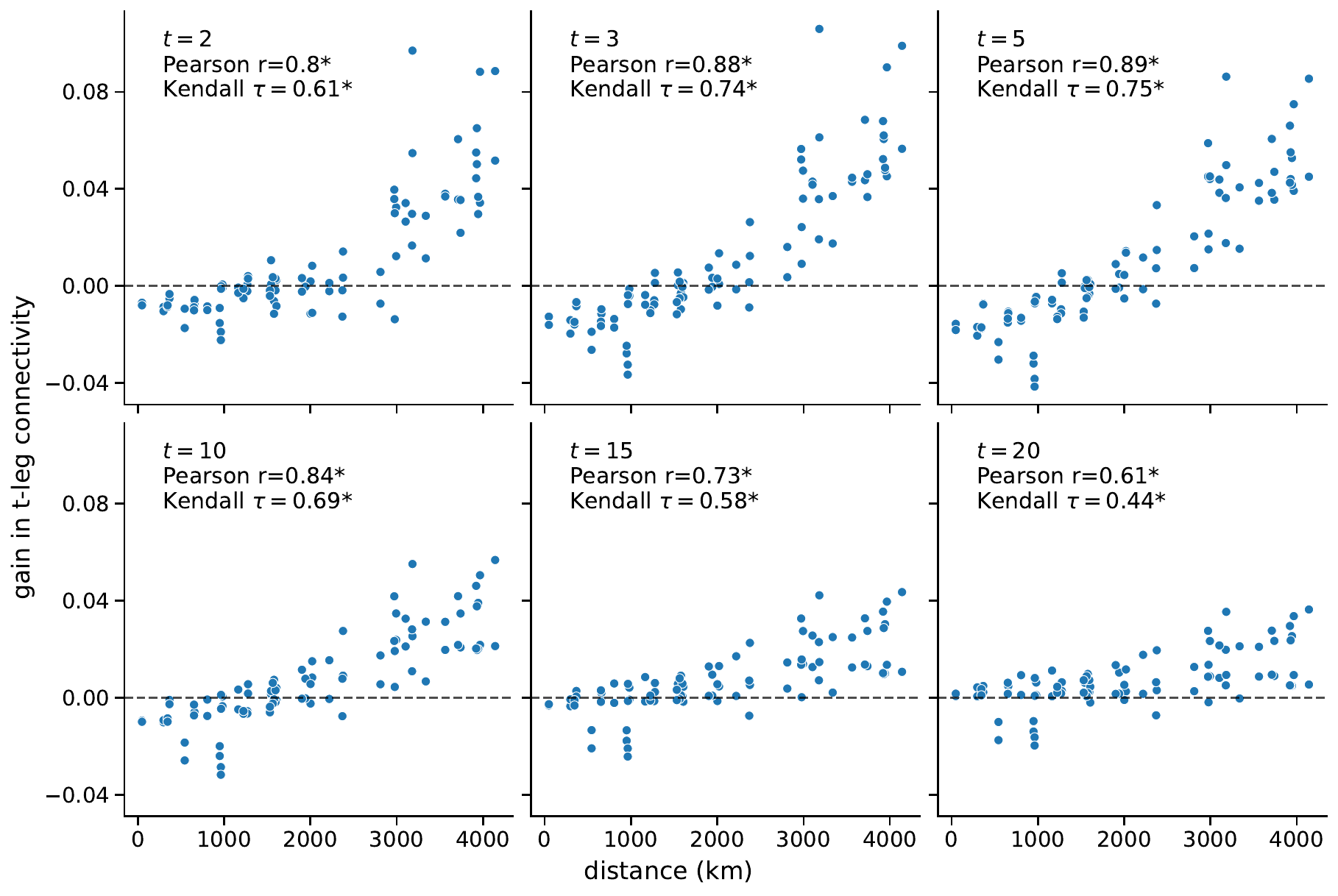}
    \caption{\textbf{Gain in connectivity.} The gain in $t$-leg connectivity (y-axis) for $t \in [2,20]$ against the distance between $o$ and $d$ (x-axis) for $(o, d)$ pairs of airports ranked 11-20 by number of transits. The black dashed line denotes no gain. $^*$ : p-value$< 10^{-8}$.}    \label{sifig:flights_connectivity_bias_sweep}
\end{figure}

\subsection{Connectivity analysis}
Generalising the notation in the main text, we define $\rho_\text{fo}(o,d,t)$ as the $t$-leg connectivity from $o$ to $d$ on $G_\text{fo}$.
\begin{equation}
     \rho_\text{fo}(o,d,t) := \left( \left(\mathbf{T^d_\text{fo}} \right)^t \right)_{od},
\end{equation}
where $\mathbf{T^d_\text{fo}}$ is $\mathbf{T_\text{fo}}$ modified to make $d$ an absorbing state. SI Fig.~\ref{sifig:flights_connectivity_bias_sweep} shows that the analysis in the main text is robust to $t$.

\section{Information flow network}
\begin{table}
\centering
\begin{tabular}{c|cc|c}
           & Corporate & Litigation & Total \\ \hline
Boston     & 19        & 29         & 48    \\
Hartford   & 8         & 11         & 19    \\
Providence & 3         & 1          & 4     \\ \hline
Total      & 30        & 41         & 71   
\end{tabular}
\caption{\textbf{Metadata in the Lazega dataset.} Cross-tabulation of office location (rows) and practice (columns) of the 71 lawyers in the data.}
\label{sitab:lazega_metadata_crosstab}
\end{table}

\subsection{Community detection}\label{sisubsec:lazega_community_detection}
The Map Equation\cite{rosvall2008maps} is a function that maps a partition of nodes to the description length of a random walk on the network using a coding scheme that leverages group structure. It takes low values for partitions where the the random walker tends to remain within modules before spreading to the rest of the network. Infomap\cite{mapequation2024software} is a greedy algorithm to optimise the Map Equation. Infomap works at the level of state nodes, allowing (1) physical nodes to be in multiple communities and (2) communities to overlap. 

\begin{figure}
    \centering
    \includegraphics[width=\linewidth]{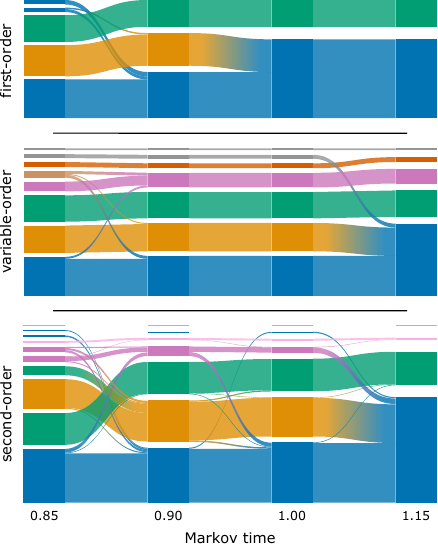}
    \caption{\textbf{Community structures across Markov times.} The changes in the community structure of $G_\text{fo}$ (top), $G_\text{c}$ (middle) and $G_\text{so}$ (bottom) networks for Markov time $\in [0.85, 1.15]$. Each block represents a module, with the colours capturing similarity. The width of the blocks and flow lines indicate the number of (state) nodes. The figures are generated using the tool at \url{https://www.mapequation.org/alluvial/}.}
    \label{sifig:lazega_alluvial_mt_sweep}
\end{figure}

\begin{table*}[]
\centering
\begin{tabular}{c|cccccc|c}
   & \multicolumn{2}{c}{Boston} & \multicolumn{2}{c}{Hartford} & \multicolumn{2}{c}{Providence} &  \\
 Module & Corp.   & Lit.   & Corp.    & Lit.    & Corp.     & Lit.     &      Total \\\hline
1        & 1           & 29           & 0            & 0             & 0             & 1              & 31    \\
2        & 18          & 0            & 1            & 0             & 3             & 0              & 22    \\
3        & 0           & 0            & 7            & 11            & 0             & 0              & 18    \\ \hline
Total    & 19          & 29           & 8            & 11            & 3             & 1              & 71   
\end{tabular}
\caption{\textbf{Community structure and metadata in $G_\text{fo}$.} Cross-tabulation of the module (rows) and metadata on office location and practice (columns) of the nodes in the first-order network. Corp. and lit. stand for corporate and litigation respectively.}
\label{sitab:lazega_fo_module_metadata_crosstab}
\end{table*}

\subsection{Picking Markov time}
The Markov time parameter can be used to find group structure across scales \cite{kheirkhahzadeh2016efficient} -- with lower values creating finer modules. To pick a scale to focus our analysis on, we explore the community structures for Markov time $\in [0.8, 1.15]$ (SI Fig.~\ref{sifig:lazega_alluvial_mt_sweep}). As Markov time increases, the major change for all three networks is the merging of the blue and orange modules into a single blue one, which happens at different Markov times for each network. These modules correlate with the metadata. The merged blue module contains the lawyers in the Boston and Providence offices, who split into corporate lawyers (blue) and litigators (orange) when we look for finer communities. We tabulate this for $G_\text{c}$ in Table \ref{sitab:lazega_fo_module_metadata_crosstab}. Our focus rests mainly on the novelty of the pink and red modules of $G_\text{c}$, which (1) persist across Markov times and (2) have no equivalent in $G_\text{fo}$. Thus, the choice of Markov time does not qualitatively affect our analysis. We pick $0.9$ since the finer communities provide insight into how individuals are grouped by work roles, and how friendship groups bridge these divisions.

\subsection{The communities of $G_\text{c}$} \label{sisubsec:comms_of_gvo}
\begin{figure}
    \centering
    \includegraphics[width=\linewidth]{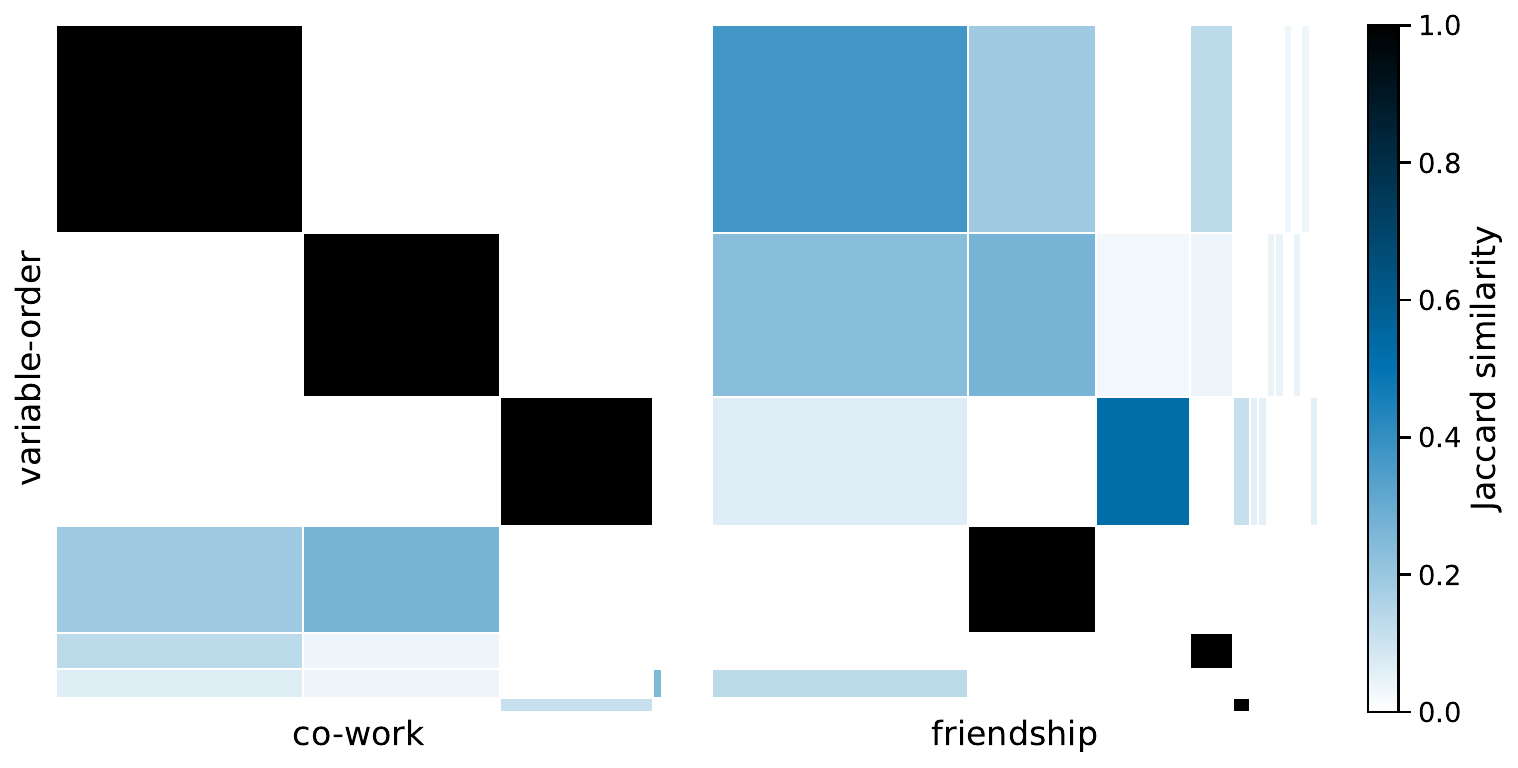}
    \caption{\textbf{Origins of the communities of $G_\text{c}$} Jaccard similarity the communities of $G_\text{c}$ (y-axis) with those of $G_\text{w}$ at Markov time 1 (x-axis, left) and $G_\text{f}$ at Markov time 1.2 a (x-axis, right). The height (resp. width) of each cell is proportional to the size of the community in $G_\text{c}$ (resp. $G_\text{w}$ and $G_\text{f}$).}
    \label{sifig:lazega_vo_wf_jaccard}
\end{figure}
To understand the origins of the community structure of $G_\text{c}$, we compare it to that of the co-work and friendship networks $G_\text{w}$ and $G_\text{f}$. We find communities corresponding to those of $G_\text{c}$ in both networks, albeit for different Markov times (SI Fig.~\ref{sifig:lazega_vo_wf_jaccard}).

%TC:endignore
\end{document}